\documentclass[journal]{IEEEtran}
\usepackage{amsmath}
\usepackage{amsfonts}
\usepackage{amssymb}
\usepackage{amstext}
\usepackage{amsthm}
\usepackage{gensymb}

\newtheorem{remark}{Remark}
\usepackage{graphicx}
\newcounter{RomanNumber}
\newcommand{\MyRoman}[1]{\setcounter{RomanNumber}{#1}\Roman{RomanNumber}}
\usepackage{algorithm}
\usepackage{algorithmic}
\usepackage{color}
\usepackage{makecell}

\begin{document}

\title{EM and SAGE algorithms for DOA estimation in the presence of unknown uniform noise}

\author{Ming-yan~Gong and Bin Lyu

\thanks{}
\thanks{M. Gong is with the School of Information and Electronics, Beijing Institute of Technology, Beijing 100081, China (e-mail: jinyan0\_o@outlook.com).}
\thanks{Bin Lyu is with the Key Laboratory of Ministry of Education in Broadband Wireless Communication and Sensor Network Technology, Nanjing University of Posts and Telecommunications, Nanjing 210003, China (e-mail: blyu@njupt.edu.cn).}
}

\markboth{} {}

\maketitle

\begin{abstract}
The expectation-maximization (EM) and space-alternating generalized EM (SAGE) algorithms have been applied to direction of arrival (DOA) estimation in known noise. In this work, the two algorithms are proposed for DOA estimation in unknown uniform noise. Both the deterministic and stochastic signal models are considered. Moreover, a modified EM (MEM) algorithm applicable to the noise assumption is also proposed. These proposed algorithms are improved to ensure the stability when the powers of sources are unequal. After being improved, numerical results illustrate that the EM algorithm has similar convergence with the MEM algorithm and the SAGE algorithm outperforms the EM and MEM algorithms for the deterministic signal model. Furthermore, numerical results show that processing the same samples from the stochastic signal model, the SAGE algorithm for the deterministic signal model requires the fewest iterations.


\end{abstract}


\IEEEpeerreviewmaketitle

\section{Introduction}

Direction of arrival (DOA) estimation is an important part of array signal processing and some high resolution estimation techniques have been proposed. In particular, the maximum-likelihood (ML) technique plays a critical role due to its superior performance. However, ML direction finding problems are non-convex and difficult to obtain their closed-form solutions, thus leading to various iterative methods of solution \cite{bibitem1}, \cite{bibitem2}.

One computationally efficient method to compute ML estimators is the expectation-maximization (EM) algorithm in \cite{bibitem3}. \cite{bibitem4} and \cite{bibitem5} have employed the EM algorithm to solve ML direction finding problems. The EM algorithm consists of two sequential steps at every iteration \cite{bibitem4}, \cite{bibitem5}: an expectation step (E-step) is to estimate the complete-data sufficient statistics by finding their conditional expectations, and a maximization step (M-step) is to estimate the signal parameters by parallel maximizations. However, the EM algorithm updates all of the parameter estimates simultaneously, which results in slow convergence. In order to speed up the convergence of the EM algorithm, \cite{bibitem6} proposes the space-alternating generalized EM (SAGE) algorithm. \cite{bibitem7} and \cite{bibitem8} show that the SAGE algorithm does yield faster convergence in terms of DOA estimation.

The EM and SAGE algorithms in ML direction finding are usually derived under known noise \cite{bibitem4}, \cite{bibitem5}, \cite{bibitem7}, \cite{bibitem8}. Know noise is with a known statistical model and without unknown parameters, which may be unrealistic in certain applications. In fact, many seminal works in ML direction finding consider the so-called unknown uniform noise model \cite{bibitem1}, \cite{bibitem2}, \cite{bibitem9}--\cite{bibitem11}. The covariance matrix of unknown uniform noise can be expressed as $\sigma\mathbf{I}_N$, where $\sigma$ is an unknown common variance and $\mathbf{I}_N$ is the $N$-order identity matrix. Under this noise assumption, \cite{bibitem9} presents a computationally attractive alternating projection algorithm for computing the deterministic ML estimator, \cite{bibitem10} investigates the statistical performance of this ML estimator and derives the Cramer-Rao lower bound, \cite{bibitem11} compares some statistical properties of both the deterministic and stochastic ML estimators. In addition to uniform noise, nonuniform noise has also attracted increasing attention. Nonuniform noise has an arbitrary diagonal covariance matrix and thus makes associated ML direction finding problems complex. For efficiently computing the deterministic and stochastic ML estimators in unknown nonuniform noise, \cite{bibitem12} and \cite{bibitem13} have presented two alternating optimization algorithms, respectively.

In this work, we develop the EM and SAGE algorithms in ML direction finding for unknown uniform noise. Theoretical analyses indicate that $\sigma$ has little effect on the two algorithms for the deterministic signal model. However, the M-step in the EM algorithm for the stochastic signal model can be no longer simplified to parallel subproblems easily when $\sigma$ is unknown. Hence, we divide the M-step into two conditional maximization steps (CM-steps) based on the expectation-CM (ECM) algorithm \cite{bibitem14}. Besides, we propose a modified EM (MEM) algorithm applicable to unknown uniform noise. Note that although the EM algorithm in \cite{bibitem15} is similar to the MEM algorithm, it is incorrectly derived.

Existing simulations about the EM and SAGE algorithms always adopt sources of equal power \cite{bibitem4}, \cite{bibitem5}, \cite{bibitem7}, \cite{bibitem8}, \cite{bibitem15}. However, we find that when the powers of sources are unequal, multiple DOA estimates obtained by the EM, MEM, and SAGE algorithms tend to be consistent with the true DOA of the source with the largest power. Hence, we improve these algorithms. 
After being improved, numerical results illustrate that 1) the EM algorithm has similar convergence with the MEM algorithm, 2) the SAGE algorithm outperforms the EM and MEM algorithms for the deterministic signal model, i.e., the SAGE algorithm converges faster and is more efficient for avoiding the convergence to an unwanted stationary point of the log-likelihood function, and 3) the SAGE algorithm cannot always outperform the EM and MEM algorithms for the stochastic signal model.

The EM, MEM, and SAGE algorithms for the deterministic signal model can process samples from the stochastic signal model. Hence, we via simulation compare the convergence of the EM and SAGE algorithms for both models. Numerical results illustrate that under the same samples, initial DOA estimates, and stopping criterion, the SAGE algorithm for the deterministic signal model requires the fewest iterations.

The contributions of this work are summarized as follows:
\begin{itemize}
\item We develop the EM and SAGE algorithms in ML direction finding for unknown uniform noise. In particular, we derive the SAGE algorithm for the stochastic signal model, which is not discussed in \cite{bibitem6}, \cite{bibitem7}, and \cite{bibitem15}.
\item We propose an MEM algorithm applicable to the unknown uniform noise assumption.
\item We improve the EM, MEM, and SAGE algorithms to ensure the stability when the powers of sources are unequal.
\item We via simulation show that the EM algorithm has similar convergence with the MEM algorithm and the SAGE algorithm outperforms the EM and MEM algorithms for the deterministic signal model. However, the SAGE algorithm cannot always outperform the EM and MEM algorithms for the stochastic signal model.
\item We via simulation show that processing the same samples from the stochastic signal model, the SAGE algorithm for the deterministic signal model requires the fewest iterations.
\end{itemize}

\emph{Notations:}
$\mathbf{a}^T$, $\mathbf{a}^H$, and $\Vert\mathbf{a}\Vert$ are the transposition, conjugate transposition, and Euclidean norm of a vector $\mathbf{a}$, respectively. $\mathbf{0}=[0~\cdots~0]^T$ and $\mathbf{1}=[1~\cdots~1]^T$. $\mathbf{A}^{-1}$, $\vert\mathbf{A}\vert$, and $\mathrm{Tr}\{\mathbf{A}\}$ are the inversion, determinant, and trace of a square matrix $\mathbf{A}$, respectively. $\mathbf{0}_N$ is the $N$-order zero matrix. $\mathbf{A}\ge\mathbf{0}_N$ and $\mathbf{A}>\mathbf{0}_N$ denote that the $N$-order square matrix $\mathbf{A}$ is positive semi-definite and definite, respectively. $\mathbb{E}\{\cdot\}$ and $\mathbb{D}\{\cdot\}$ denote expectation and variance, respectively. $\jmath$ is the imaginary unit.

\section{Data Model and Problem Formulation}

We consider an array composed of $N$ isotropic sensors receiving the signals emitted by $M$ narrowband far-field sources with the same known center wavelength $\lambda$. The array geometry is arbitrary and known in a Cartesian coordinate system and let the $n$th sensor be at a known position $\mathbf{p}_{n} = [x_n~y_n~z_n]^{T}$ close to the origin. Spherical coordinates are used to show the DOAs of the $M$ sources. The elevation and azimuth angles of the $m$th source are denoted by $\varphi_m\in [0, \pi]$ and $\phi_m\in [0, 2\pi)$, respectively. Thus, its unit directional vector can be expressed as $\mathbf{q}_m=[\sin(\varphi_m)\cos(\phi_m)~\sin(\varphi_m)\sin(\phi_m)~\cos(\varphi_m)]^{T}$ in the Cartesian coordinate system.

We use the origin as the reference point of the array. Then, the phase difference, with respect to the $m$th source, between the two signals, respectively, received at the origin and the $n$th sensor can be approximated as $\psi_{n,m} = -\frac{2\pi}{\lambda}\mathbf{p}_{n}^{T}\mathbf{q}_m$. After being down-converted to baseband, the received signal vector of the array can be characterized by \cite{bibitem1}, \cite{bibitem2}
\begin{eqnarray}     
\mathbf{y}(t)={\sum}_{m=1}^M\mathbf{a}(\theta_m)s_m(t)+\mathbf{w}(t),
\end{eqnarray}
where $\theta_m=[\varphi_m~\phi_m]^T\in\Omega$, $\mathbf{a}(\theta_m)=[e^{\jmath\psi_{1,m}}~\cdots~e^{\jmath\psi_{N,m}}]^T$,
$s_m(t)$ is the $m$th source baseband signal received at the origin and its power is $P_m$, $\mathbf{w}(t)\sim\mathcal{CN}(\textbf{0},\sigma\mathbf{I}_N)$ represents a white Gaussian noise vector with a common noise variance $\sigma$.

EM-type algorithms require the definition of the underlying complete data and their associated log-likelihood functions. According to the EM paradigm in \cite{bibitem4} and \cite{bibitem5}, after sampling we design the samples (``snapshots") or incomplete data as
\begin{eqnarray}     
\mathbf{y}(t)&=&{\sum}_{m=1}^M\big[\mathbf{a}(\theta_m)s_m(t)+\mathbf{w}_m(t)\big]\nonumber \\
&=&{\sum}_{m=1}^M\mathbf{x}_m(t), t=1,2,\dots,T,
\end{eqnarray}
where $\mathbf{w}_m(t)\sim\mathcal{CN}(\textbf{0},\alpha_m\sigma\mathbf{I}_N)$ and the $\mathbf{w}_m(t)$'s are mutually uncorrelated, $\boldsymbol{\alpha}=[\alpha_1~\cdots~\alpha_M]^T\ge\mathbf{0}$ and $\mathbf{1}^T\boldsymbol{\alpha}=1$, the $\mathbf{x}_m(t)$'s are the complete data, $T$ is the number of samples.

Since the joint probability density functions (PDFs) of the incomplete and complete data depend also on the statistical model of the $s_m(t)$'s, we consider the deterministic and stochastic signal models separately.

\begin{subsection}{Deterministic Signal Model}

The deterministic signal model requires the deterministic, arbitrary, and unknown $s_m(t)$'s \cite{bibitem4}, \cite{bibitem5}, \cite{bibitem9}--\cite{bibitem11}, i.e.,
$\mathbf{x}_m(t)\sim\mathcal{CN}(\mathbf{a}(\theta_m)s_m(t),\alpha_m\sigma\mathbf{I}_N)$,
$\mathbf{y}(t)\sim\mathcal{CN}(\mathbf{A}(\boldsymbol{\theta})\mathbf{s}(t),\sigma\mathbf{I}_N)$ with $\mathbf{A}(\boldsymbol{\theta})=[\mathbf{a}(\theta_1)~\cdots~\mathbf{a}(\theta_M)]$, $\boldsymbol{\theta}=[\theta_1~\cdots~\theta_M]\in\boldsymbol{\Omega}~(\boldsymbol{\Omega}=\Omega^M)$, and $\mathbf{s}(t)=[s_1(t)~\cdots~s_M(t)]^T$. Then, the incomplete- and complete-data log-likelihood functions are, respectively, expressed as
\begin{subequations}
\begin{eqnarray}       
&\mathcal{L}(\boldsymbol{\Theta},\sigma)=\ln p(\mathbf{Y};\boldsymbol{\theta},\mathbf{S},\sigma)\nonumber\\
=&-TN\ln(\pi\sigma)-\frac{1}{\sigma}\sum_{t=1}^T\Vert\mathbf{y}(t)-\mathbf{A}(\boldsymbol{\theta})\mathbf{s}(t)\Vert^2,
\end{eqnarray}
\begin{eqnarray}       
&\mathcal{Q}(\mathbf{X}_1,\dots,\mathbf{X}_M,\boldsymbol{\Theta},\sigma)
=\ln p(\mathbf{X}_1,\dots,\mathbf{X}_M;\boldsymbol{\theta},\mathbf{S},\sigma)\nonumber\\
=&-TMN\ln(\pi\sigma)-TN\sum_{m=1}^M\ln(\alpha_m)-\nonumber\\
&\frac{1}{\sigma}\sum_{t=1}^T\sum_{m=1}^M\frac{1}{\alpha_m}\Vert\mathbf{x}_m(t)-\mathbf{a}(\theta_m)s_m(t)\Vert^2,
\end{eqnarray}
\end{subequations}
where $\mathbf{S}=[\mathbf{s}(1)~\cdots~\mathbf{s}(T)]$, $\mathbf{Y}=[\mathbf{y}(1)~\cdots~\mathbf{y}(T)]$, $\mathbf{X}_m=[\mathbf{x}_m(1)~\cdots~\mathbf{x}_m(T)]$, and $\boldsymbol{\Theta}=(\boldsymbol{\theta},\mathbf{S})$ denotes the signal parameters while $\sigma$ is the only noise parameter. Thus, the ML estimation problem, i.e., $\max_{\boldsymbol{\theta}\in\boldsymbol{\Omega},\mathbf{S},\sigma>0}\mathcal{L}(\boldsymbol{\Theta},\sigma)$, can be simplified to
\begin{eqnarray}       
\mathop{\min_{\boldsymbol{\theta}\in\boldsymbol{\Omega},\mathbf{S},\sigma>0}}N\ln(\sigma)+
\frac{1}{\sigma T}\sum_{t=1}^T\big\Vert\mathbf{y}(t)-\mathbf{A}(\boldsymbol{\theta})\mathbf{s}(t)\big\Vert^2.
\end{eqnarray}

\end{subsection}

\begin{subsection}{Stochastic Signal Model}
In the stochastic signal model, $s_m(t)\sim\mathcal{CN}(0,P_m)$. For simplicity, all of the $s_m(t)$'s and $\mathbf{w}_m(t)$'s are assumed to be mutually uncorrelated \cite{bibitem4}, \cite{bibitem5}, i.e., $\mathbf{x}_m(t)\sim\mathcal{CN}(\textbf{0},\mathbf{C}_m)$ with $\mathbf{C}_m=P_m\mathbf{a}(\theta_m)\mathbf{a}^H(\theta_m)+\alpha_m\sigma\mathbf{I}_N$, and
$\mathbf{y}(t)\sim\mathcal{CN}(\mathbf{0},\mathbf{C}_y)$ with $\mathbf{C}_y=\sum_{m=1}^M\mathbf{C}_m$. Then, the incomplete- and complete-data log-likelihood functions are, respectively, expressed as
\begin{subequations}
\begin{eqnarray}       
&\mathcal{L}(\boldsymbol{\Theta},\sigma)=\ln p(\mathbf{Y};\boldsymbol{\theta},\mathbf{P},\sigma)\nonumber \\
=&-TN\ln(\pi)-T\ln\vert\mathbf{C}_y\vert-\sum_{t=1}^T\mathbf{y}^H(t)\mathbf{C}_y^{-1}\mathbf{y}(t),
\end{eqnarray}
\begin{eqnarray}       
&\mathcal{Q}(\mathbf{X}_1,\dots,\mathbf{X}_M,\boldsymbol{\Theta},\sigma)=\ln p(\mathbf{X}_1,\dots,\mathbf{X}_M;\boldsymbol{\theta},\mathbf{P},\sigma)\nonumber \\
=&-TMN\ln(\pi)-T\sum_{m=1}^M\ln\vert\mathbf{C}_m\vert-\nonumber \\
&\sum_{t=1}^T\sum_{m=1}^M\mathbf{x}^H_m(t)\mathbf{C}_m^{-1}\mathbf{x}_m(t),
\end{eqnarray}
\end{subequations}
where $\mathbf{P}=[P_1~\cdots~P_M]^T$ and $\boldsymbol{\Theta}=(\boldsymbol{\theta},\mathbf{P})$. Finally, the ML estimation problem can be simplified to
\begin{eqnarray}       
\min_{\boldsymbol{\theta}\in\boldsymbol{\Omega},\mathbf{P}\ge\mathbf{0},\sigma>0}
\ln\vert\mathbf{C}_y\vert+\mathrm{Tr}\big\{\mathbf{C}_y^{-1}\hat{\mathbf{R}}_y\big\},
\end{eqnarray}
where $\hat{\mathbf{R}}_y=\frac{1}{T}\sum_{t=1}^T\mathbf{y}(t)\mathbf{y}^H(t)\ge\mathbf{0}_N$ is the sample covariance matrix.

\end{subsection}

\section{EM Algorithm}

In this section, we derive the EM algorithm for solving problems (4) and (6). For convenience, we define the following notations:
\begin{itemize}
\item[1)]$(\cdot)^{(k)}$ denotes an iterative value obtained during the $k$th iteration and $(\cdot)^{(0)}$ denotes an initial value.
\item[2)]$\hat{\mathbf{R}}_m=\frac{1}{T}\sum_{t=1}^T\mathbf{x}_m(t)\mathbf{x}^H_m(t)\ge\mathbf{0}_N$.
\item[3)]$\boldsymbol{\Pi}_m=\frac{1}{N}\mathbf{a}(\theta_m)\mathbf{a}^H(\theta_m)$.
\item[4)]$e^{(k)}_m=\mathrm{Tr}\big\{\boldsymbol{\Pi}^{(k)}_m\hat{\mathbf{R}}^{(k)}_m\big\}\ge0$.
\item[5)]$d^{(k)}_m=\mathrm{Tr}\big\{\big(\mathbf{I}_N-\boldsymbol{\Pi}^{(k)}_m\big)\hat{\mathbf{R}}^{(k)}_m\big\}
=\mathrm{Tr}\big\{\hat{\mathbf{R}}^{(k)}_m\big\}-e^{(k)}_m\ge0$.
\end{itemize}

\subsection{Deterministic Signal Model}

At the $k$th E-step, the conditional expectation of (3b) is computed by
\begin{eqnarray}       
&\mathbb{E}\left\{\mathcal{Q}(\mathbf{X}_1,\dots,\mathbf{X}_M,\boldsymbol{\Theta},\sigma)|\mathbf{Y};\boldsymbol{\Theta}^{(k-1)},\sigma^{(k-1)}\right\}\nonumber\\
=&-T\Big\{MN\ln(\pi\sigma)+N\sum_{m=1}^M\ln(\alpha_m)+\frac{1}{\sigma}\big[c^{(k)}+\nonumber\\
&\frac{1}{T}\sum_{t=1}^T\sum_{m=1}^M\frac{1}{\alpha_m}\Vert\mathbf{x}^{(k)}_m(t)-\mathbf{a}(\theta_m)s_m(t)\Vert^2\big]\Big\},
\end{eqnarray}
where the conditional PDF of $\mathbf{x}_m(t)$ can be derived from \cite{bibitem17} and
\begin{subequations}
\begin{eqnarray}       
\mathbf{x}^{(k)}_m(t)=&\mathbb{E}\left\{\mathbf{x}_m(t)|\mathbf{Y};\boldsymbol{\Theta}^{(k-1)},\sigma^{(k-1)}\right\}\nonumber\\
=&\alpha_m\big[\mathbf{y}(t)-\mathbf{A}(\boldsymbol{\theta}^{(k-1)})\mathbf{s}^{(k-1)}(t)\big]\nonumber\\
&+\mathbf{a}(\theta^{(k-1)}_m)s^{(k-1)}_m(t),\forall m,t,
\end{eqnarray}
\begin{eqnarray}       
c^{(k)}&=&\frac{1}{T}\sum_{t=1}^T\sum_{m=1}^M\frac{1}{\alpha_m}\mathrm{Tr}\left\{\mathbb{D}\big\{\mathbf{x}_m(t)|\mathbf{Y};\boldsymbol{\Theta}^{(k-1)},\sigma^{(k-1)}\big\}\right\}\nonumber\\
&=&\frac{1}{T}\sum_{t=1}^T\sum_{m=1}^M\frac{1}{\alpha_m}\mathrm{Tr}\left\{\alpha_m(1-\alpha_m)\sigma^{(k-1)}\mathbf{I}_N\right\}\nonumber\\
&=&N(M-1)\sigma^{(k-1)}.
\end{eqnarray}
\end{subequations}

At the $k$th M-step, based on (7) the EM algorithm updates the estimates of $\boldsymbol{\Theta}$ and $\sigma$ by
\begin{eqnarray}       
\min_{\boldsymbol{\theta}\in\boldsymbol{\Omega},\mathbf{S},\sigma>0}&MN\ln(\sigma)+\frac{1}{\sigma}\big[c^{(k)}+\nonumber\\
&\frac{1}{T}\sum_{t=1}^T\sum_{m=1}^M\frac{1}{\alpha_m}\Vert\mathbf{x}^{(k)}_m(t)-\mathbf{a}(\theta_m)s_m(t)\Vert^2\big],
\end{eqnarray}
which can be solved easily in a separable fashion. Then, we have the parameter estimates:
\begin{subequations}
\begin{eqnarray}       
\theta^{(k)}_m=\arg\max_{\theta_m\in\Omega}\mathrm{Tr}\big\{\boldsymbol{\Pi}_m\hat{\mathbf{R}}^{(k)}_m\big\},\forall m,
\end{eqnarray}
\begin{eqnarray}       
s^{(k)}_m(t)=\frac{1}{N}\mathbf{a}^H(\theta^{(k)}_m)\mathbf{x}^{(k)}_m(t),\forall m,t,
\end{eqnarray}
\begin{eqnarray}       
\sigma^{(k)}=(1-\frac{1}{M})\sigma^{(k-1)}+\frac{1}{MN}\sum_{m=1}^Md^{(k)}_m/\alpha_m,
\end{eqnarray}
\end{subequations}
where $\hat{\mathbf{R}}^{(k)}_m=\frac{1}{T}\sum_{t=1}^T\mathbf{x}^{(k)}_m(t)[\mathbf{x}^{(k)}_m(t)]^H$ and $\sigma^{(k)}>0$ if $\sigma^{(k-1)}>0$.

\begin{remark}
Eliminating $\sigma$, problem (4) can be simplified to
\begin{eqnarray}       
\min_{\boldsymbol{\theta}\in\boldsymbol{\Omega},\mathbf{S}}\frac{1}{T}\sum_{t=1}^T\big\Vert\mathbf{y}(t)-\mathbf{A}(\boldsymbol{\theta})\mathbf{s}(t)\big\Vert^2,
\end{eqnarray}
which indicates that for the deterministic signal model, the ML estimator of $\boldsymbol{\Theta}$ is unrelated to $\sigma$. Additionally, note that the $\mathbf{x}^{(k)}_m(t)$'s in (8a), the $\theta^{(k)}_m$'s in (10a), and the $s^{(k)}_m(t)$'s in (10b) are unrelated to $\sigma^{(k-1)}$ and respectively identical with these in the EM algorithm under the known $\sigma$ \cite{bibitem5}, \cite{bibitem8}. Then, we can conclude that the EM algorithm under the unknown $\sigma$ is equivalent to that under the known $\sigma$ when not estimating $\sigma$. Accordingly, (10c) can be omitted when the algorithm does not consider the nuisance parameter $\sigma$.
\end{remark}

\subsection{Stochastic Signal Model}

Note that $\hat{\mathbf{R}}_m$ is a complete-data sufficient statistic for $\mathbf{C}_m$ \cite{bibitem5}, \cite{bibitem16}, which contains $\theta_m$, $P_m$, and $\sigma$. At the $k$th E-step, the EM algorithm estimates the $\hat{\mathbf{R}}_m$'s by finding their conditional expectations \cite{bibitem3}:
\begin{eqnarray}       
\hat{\mathbf{R}}^{(k)}_m=&\mathbb{E}\left\{\hat{\mathbf{R}}_m|\mathbf{Y};\boldsymbol{\Theta}^{(k-1)},\sigma^{(k-1)}\right\}\nonumber\\
=&\mathbf{C}^{(k-1)}_m[\mathbf{C}^{(k-1)}_y]^{-1}\hat{\mathbf{R}}_y[\mathbf{C}^{(k-1)}_y]^{-1}\mathbf{C}^{(k-1)}_m+\nonumber\\
&\left\{\mathbf{C}^{(k-1)}_m-\mathbf{C}^{(k-1)}_m[\mathbf{C}^{(k-1)}_y]^{-1}\mathbf{C}^{(k-1)}_m\right\},\forall m,
\end{eqnarray}
where
\begin{eqnarray}
&\mathbb{E}\left\{\mathbf{x}_m(t)|\mathbf{Y};\boldsymbol{\Theta}^{(k-1)},\sigma^{(k-1)}\right\}\nonumber\\
=&\mathbf{C}^{(k-1)}_m[\mathbf{C}^{(k-1)}_y]^{-1}\mathbf{y}(t),\forall m,t,\nonumber\\
&\mathbb{D}\left\{\mathbf{x}_m(t)|\mathbf{Y};\boldsymbol{\Theta}^{(k-1)},\sigma^{(k-1)}\right\}\nonumber\\
=&\mathbf{C}^{(k-1)}_m-\mathbf{C}^{(k-1)}_m[\mathbf{C}^{(k-1)}_y]^{-1}\mathbf{C}^{(k-1)}_m\ge\mathbf{0}_N,\forall m,t.\nonumber
\end{eqnarray}

At the $k$th M-step, based on (5b) and (12) the EM algorithm updates the estimates of $\boldsymbol{\Theta}$ and $\sigma$ by
\begin{eqnarray} 
\min_{\boldsymbol{\theta}\in\boldsymbol{\Omega},\mathbf{r}\ge\mathbf{0},\sigma>0}&MN\ln(\sigma)+\sum_{m=1}^M\big[\ln\vert\mathbf{D}_m\vert\nonumber\\
&+\frac{1}{\sigma}\mathrm{Tr}\big\{\mathbf{D}_m^{-1}\hat{\mathbf{R}}^{(k)}_m\big\}\big],
\end{eqnarray}
where $\mathbf{C}_m=\sigma\mathbf{D}_m$ with $\mathbf{D}_m=r_m\mathbf{a}(\theta_m)\mathbf{a}^H(\theta_m)+\alpha_m\mathbf{I}_N$ and $\mathbf{r}=[r_1~\cdots~r_M]^T$ with $r_m=\frac{P_m}{\sigma}$ being the signal to noise ratio of the $m$th source.

Due to $\sigma$, it seems to be very difficult that problem (13) can be simplified to parallel subproblems. To perform the M-step simply, we divide it into two CM-steps based on the ECM algorithm \cite{bibitem14}.
\begin{itemize}
\item \emph{First CM-step:} estimate $\boldsymbol{\Theta}$ but hold $\sigma=\sigma^{(k-1)}$ fixed. Then, (13) can be simplified to the $M$ parallel subproblems
\begin{eqnarray} 
\min_{\theta_m\in\Omega,r_m\ge0}\ln\vert\mathbf{D}_m\vert+\frac{1}{\sigma^{(k-1)}}\mathrm{Tr}\big\{\mathbf{D}_m^{-1}\hat{\mathbf{R}}^{(k)}_m\big\},\forall m,
\end{eqnarray}
which can be solved using the method in \cite{bibitem5}. Accordingly, the estimate of $\boldsymbol{\Theta}$ is updated by
\begin{subequations}
\begin{eqnarray}     
\theta^{(k)}_m=
\arg\max_{\theta_m\in\Omega}\mathrm{Tr}\big\{\boldsymbol{\Pi}_m\hat{\mathbf{R}}^{(k)}_m\big\}, \forall m,
\end{eqnarray}
\begin{eqnarray}       
P^{(k)}_m=&r^{(k)}_m\sigma^{(k-1)}\nonumber\\
=&\max\left\{\frac{1}{N}\big(e^{(k)}_m-\alpha_m\sigma^{(k-1)}\big),0\right\},\forall m,
\end{eqnarray}
\end{subequations}
where $\theta^{(k)}_m$ is indeterminate if $P^{(k)}_m=0$.
\item \emph{Second CM-step:} estimate $\sigma$ but hold $\boldsymbol{\Theta}=\boldsymbol{\Theta}^{(k)}$ fixed. Then, (13) can be simplified to
\begin{eqnarray} 
\min_{\sigma>0}MN\ln(\sigma)+\frac{1}{\sigma}\sum_{m=1}^M\mathrm{Tr}\left\{[\mathbf{D}^{(k)}_m]^{-1}\hat{\mathbf{R}}^{(k)}_m\right\}.
\end{eqnarray}
Thus, the estimate of $\sigma$ is updated by
\begin{eqnarray} 
\sigma^{(k)}&=&\frac{1}{MN}\sum_{m=1}^M\mathrm{Tr}\left\{[\mathbf{D}^{(k)}_m]^{-1}\hat{\mathbf{R}}^{(k)}_m\right\}\nonumber\\
&=&\frac{1}{N}\sigma^{(k-1)}+\frac{1}{MN}\sum_{m=1}^Md^{(k)}_m/\alpha_m,
\end{eqnarray}
where $\sigma^{(k)}>0$ if $\sigma^{(k-1)}>0$.
\end{itemize}

\begin{remark}
Although the M-step (13) of the EM algorithm is divided into the two sequential CM-steps (14) and (16), the monotonicity of the algorithm still holds \cite{bibitem14}, i.e.,
\begin{eqnarray}       
\mathcal{L}(\boldsymbol{\Theta}^{(k)},\sigma^{(k)})\ge\mathcal{L}(\boldsymbol{\Theta}^{(k-1)},\sigma^{(k-1)}).\nonumber
\end{eqnarray}
Obviously, this algorithm is only a generalized EM (GEM) algorithm \cite{bibitem3} but for convenience, we still call it the EM algorithm.
\end{remark}

\section{MEM Algorithm}

In the previous section, $\boldsymbol{\alpha}$ is fixed and known. In this section, we regard $\boldsymbol{\alpha}$ as a parameter to be estimated and thus propose an MEM algorithm applicable to the unknown uniform noise assumption.

In order to estimate $\sigma$ and $\boldsymbol{\alpha}$ in the MEM algorithm easily, we introduce $\sigma_m=\alpha_m\sigma$ as the common noise variance of the $m$th source, i.e.,
\begin{eqnarray}       
\mathbf{w}_m(t)\sim\mathcal{CN}(\mathbf{0},\sigma_m\mathbf{I}_N).
\end{eqnarray}
Then, $\sigma$ and $\boldsymbol{\alpha}$ can be estimated by $\sigma=\sum_{m=1}^M\sigma_m$ and $\alpha_m=\sigma_m/\sigma$ after estimating $\boldsymbol{\sigma}=[\sigma_1~\cdots~\sigma_M]^T$ at the M-step of the MEM algorithm. In addition, we first assume $\boldsymbol{\sigma}^{(k)}>\mathbf{0}$ for $k\ge0$.

\subsection{Deterministic Signal Model}

According to (18), (3b) is rewritten as
\begin{eqnarray}       
&\mathcal{Q}(\mathbf{X}_1,\dots,\mathbf{X}_M,\boldsymbol{\Theta},\boldsymbol{\sigma})
=\ln p(\mathbf{X}_1,\dots,\mathbf{X}_M;\boldsymbol{\theta},\mathbf{S},\boldsymbol{\sigma})\nonumber\\
=&-TMN\ln(\pi)-TN\sum_{m=1}^M\ln(\sigma_m)-\nonumber\\
&\sum_{t=1}^T\sum_{m=1}^M\frac{1}{\sigma_m}\Vert\mathbf{x}_m(t)-\mathbf{a}(\theta_m)s_m(t)\Vert^2.
\end{eqnarray}
With the help of (19), at the $k$th E-step the MEM algorithm computes the conditional expectation:
\begin{eqnarray}       
&\mathbb{E}\left\{\mathcal{Q}(\mathbf{X}_1,\dots,\mathbf{X}_M,\boldsymbol{\Theta},\boldsymbol{\sigma})|\mathbf{Y};\boldsymbol{\Theta}^{(k-1)},\boldsymbol{\sigma}^{(k-1)}\right\}\nonumber\\
=&-TMN\ln(\pi)-T\sum_{m=1}^M\Big\{N\ln(\sigma_m)+\frac{1}{\sigma_m}\big[c^{(k)}_m\nonumber\\
&+\frac{1}{T}\sum_{t=1}^T\Vert\mathbf{x}^{(k)}_m(t)-\mathbf{a}(\theta_m)s_m(t)\Vert^2\big]\Big\},
\end{eqnarray}
where
\begin{subequations}
\begin{eqnarray}       
\mathbf{x}^{(k)}_m(t)=&\mathbb{E}\left\{\mathbf{x}_m(t)|\mathbf{Y};\boldsymbol{\Theta}^{(k-1)},\boldsymbol{\sigma}^{(k-1)}\right\}\nonumber\\
=&\sigma^{(k-1)}_m/\sigma^{(k-1)}\big[\mathbf{y}(t)-\mathbf{A}(\boldsymbol{\theta}^{(k-1)})\mathbf{s}^{(k-1)}(t)\big]\nonumber\\
&+\mathbf{a}(\theta^{(k-1)}_m)s^{(k-1)}_m(t),\forall m,t,
\end{eqnarray}
\begin{eqnarray}       
c^{(k)}_m&=&\frac{1}{T}\sum_{t=1}^T\mathrm{Tr}\left\{\mathbb{D}\big\{\mathbf{x}_m(t)|\mathbf{Y};\boldsymbol{\Theta}^{(k-1)},\boldsymbol{\sigma}^{(k-1)}\big\}\right\}\nonumber\\
&=&\frac{1}{T}\sum_{t=1}^T\mathrm{Tr}\left\{\sigma^{(k-1)}_m\big(1-\sigma^{(k-1)}_m/\sigma^{(k-1)}\big)\mathbf{I}_N\right\}\nonumber\\
&=&N\sigma^{(k-1)}_m\big(1-\sigma^{(k-1)}_m/\sigma^{(k-1)}\big),\forall m.
\end{eqnarray}
\end{subequations}

At the $k$th M-step, based on (20) the MEM algorithm updates the estimates of $\boldsymbol{\Theta}$ and $\boldsymbol{\sigma}$ by the following $M$ parallel subproblems:
\begin{eqnarray}       
\min_{\theta_m\in\Omega,\mathbf{s}_m,\sigma_m>0}&N\ln(\sigma_m)+\frac{1}{\sigma_m}\big[c^{(k)}_m+\nonumber\\
&\frac{1}{T}\sum_{t=1}^T\Vert\mathbf{x}^{(k)}_m(t)-\mathbf{a}(\theta_m)s_m(t)\Vert^2\big],\forall m,
\end{eqnarray}
where $\mathbf{s}_m=[s_m(1)~\cdots~s_m(T)]$. Then, we can obtain the parameter estimates:
\begin{subequations}
\begin{eqnarray}       
\theta^{(k)}_m=\arg\max_{\theta_m\in\Omega}\mathrm{Tr}\big\{\boldsymbol{\Pi}_m\hat{\mathbf{R}}^{(k)}_m\big\},\forall m,
\end{eqnarray}
\begin{eqnarray}       
s^{(k)}_m(t)=\frac{1}{N}\mathbf{a}^H(\theta^{(k)}_m)\mathbf{x}^{(k)}_m(t),\forall m,t,
\end{eqnarray}
\begin{eqnarray}       
\sigma^{(k)}_m=\sigma^{(k-1)}_m\big(1-\sigma^{(k-1)}_m/\sigma^{(k-1)}\big)+d^{(k)}_m/N,\forall m,
\end{eqnarray}
\end{subequations}
where $\hat{\mathbf{R}}^{(k)}_m=\frac{1}{T}\sum_{t=1}^T\mathbf{x}^{(k)}_m(t)[\mathbf{x}^{(k)}_m(t)]^H$. Note that when $\boldsymbol{\sigma}^{(k-1)}>\mathbf{0}$, based on (23c) we have $1-\sigma^{(k-1)}_m/\sigma^{(k-1)}>0,\forall m$, and $\sigma^{(k)}_m>0,\forall m$, i.e., $\boldsymbol{\sigma}^{(k)}>\mathbf{0}$.

\begin{remark}
The $\mathbf{x}^{(k)}_m(t)$'s in (21a) are related to $\boldsymbol{\sigma}^{(k-1)}$, so the $\theta^{(k)}_m$'s and $s^{(k)}_m(t)$'s in (23) are related to $\boldsymbol{\sigma}^{(k-1)}$, i.e., iterative knowledge associated with $\sigma$ is utilized to estimate $\boldsymbol{\Theta}$ in the MEM algorithm for the deterministic signal model.
\end{remark}

\subsection{Stochastic Signal Model}

According to (18), $\mathbf{x}_m(t)\sim\mathcal{CN}(\mathbf{0},\mathbf{C}_m)$ with $\mathbf{C}_m=P_m\mathbf{a}(\theta_m)\mathbf{a}^H(\theta_m)+\sigma_m\mathbf{I}_N$ and (5b) is rewritten as
\begin{eqnarray}       
&\mathcal{Q}(\mathbf{X}_1,\dots,\mathbf{X}_M,\boldsymbol{\Theta},\boldsymbol{\sigma})=\ln p(\mathbf{X}_1,\dots,\mathbf{X}_M;\boldsymbol{\theta},\mathbf{P},\boldsymbol{\sigma})\nonumber \\
=&-T\sum_{m=1}^M\big[N\ln(\pi)+
\ln\vert\mathbf{C}_m\vert+\mathrm{Tr}\big\{\mathbf{C}_m^{-1}\hat{\mathbf{R}}_m\big\}\big].
\end{eqnarray}
With the help of (24), at the $k$th E-step the MEM algorithm estimates the $\hat{\mathbf{R}}_m$'s by finding their conditional expectations:
\begin{eqnarray}       
\hat{\mathbf{R}}^{(k)}_m=&\mathbb{E}\left\{\hat{\mathbf{R}}_m|\mathbf{Y};\boldsymbol{\Theta}^{(k-1)},\boldsymbol{\sigma}^{(k-1)}\right\}\nonumber\\
=&\mathbf{C}^{(k-1)}_m[\mathbf{C}^{(k-1)}_y]^{-1}\hat{\mathbf{R}}_y[\mathbf{C}^{(k-1)}_y]^{-1}\mathbf{C}^{(k-1)}_m+\nonumber\\
&\left\{\mathbf{C}^{(k-1)}_m-\mathbf{C}^{(k-1)}_m[\mathbf{C}^{(k-1)}_y]^{-1}\mathbf{C}^{(k-1)}_m\right\},\forall m,
\end{eqnarray}
where
\begin{eqnarray}       
\mathbf{C}^{(k-1)}_m[\mathbf{C}^{(k-1)}_y]^{-1}\hat{\mathbf{R}}_y[\mathbf{C}^{(k-1)}_y]^{-1}\mathbf{C}^{(k-1)}_m\ge\mathbf{0}_N,\forall m,\nonumber\\
\mathbf{C}^{(k-1)}_m-\mathbf{C}^{(k-1)}_m[\mathbf{C}^{(k-1)}_y]^{-1}\mathbf{C}^{(k-1)}_m\ge\mathbf{0}_N,\forall m.\nonumber
\end{eqnarray}

At the $k$th M-step, based on (24) and (25) the MEM algorithm estimates $\boldsymbol{\Theta}$ and $\boldsymbol{\sigma}$ by the following $M$ parallel subproblems\footnote{Unlike the M-step (13) in the EM algorithm for the stochastic signal model, the M-step in the MEM algorithm for the stochastic signal model can be easily simplified to parallel subproblems.}:
\begin{eqnarray}       
\min_{\theta_m\in\Omega,r_m\ge0,\sigma_m>0}&N\ln(\sigma_m)+\ln\vert\mathbf{D}_m\vert+\nonumber\\
&\frac{1}{\sigma_m}\mathrm{Tr}\big\{\mathbf{D}_m^{-1}\hat{\mathbf{R}}^{(k)}_m\big\},\forall m,
\end{eqnarray}
where $\mathbf{C}_m=\sigma_m\mathbf{D}_m$ with $\mathbf{D}_m=r_m\mathbf{a}(\theta_m)\mathbf{a}^H(\theta_m)+\mathbf{I}_N$ and $r_m=P_m/\sigma_m$. Since $\vert\mathbf{D}_m\vert=Nr_m+1$ and $\mathbf{D}_m^{-1}=\mathbf{I}_N-\frac{Nr_m}{Nr_m+1}\boldsymbol{\Pi}_m$, subproblems (26) can be rewritten as
\begin{eqnarray}       
\min_{\theta_m\in\Omega,r_m\ge0,\sigma_m>0}N\ln(\sigma_m)+\ln(Nr_m+1)+\nonumber\\
\frac{1}{\sigma_m}\mathrm{Tr}\big\{\hat{\mathbf{R}}^{(k)}_m\big\}-\frac{Nr_m}{\sigma_m(Nr_m+1)}\mathrm{Tr}\big\{\boldsymbol{\Pi}_m\hat{\mathbf{R}}^{(k)}_m\big\},\forall m.
\end{eqnarray}
We first simplify subproblems (27) by eliminating $\mathbf{r}$ \cite{bibitem18}, \cite{bibitem19}. Hence, after estimating $\boldsymbol{\theta}$ and $\boldsymbol{\sigma}$, $\mathbf{r}$ and $\mathbf{P}$ are estimated by
\begin{eqnarray}       
r^{(k)}_m=\max\left\{\frac{1}{N}\big(e^{(k)}_m/\sigma^{(k)}_m-1\big),0\right\},\forall m,\nonumber
\end{eqnarray}
\begin{eqnarray}       
P^{(k)}_m=r^{(k)}_m\sigma^{(k)}_m=\max\left\{\frac{1}{N}\big(e^{(k)}_m-\sigma^{(k)}_m\big),0\right\},\forall m,\nonumber
\end{eqnarray}
which imply that if $r^{(k)}_m=0$, $\theta^{(k)}_m$ will be indeterminate and $\sigma^{(k)}_m=\frac{1}{N}\mathrm{Tr}\big\{\hat{\mathbf{R}}^{(k)}_m\big\}$ by (27). However, to estimate $\boldsymbol{\theta}$ and $\boldsymbol{\sigma}$, we must assume $\mathbf{r}^{(k)}>\mathbf{0}$. After eliminating $\mathbf{r}$, subproblems (27) are simplified to
\begin{eqnarray}       
\min_{\theta_m\in\Omega,\sigma_m>0}&(N-1)\ln(\sigma_m)+\ln\left(\mathrm{Tr}\big\{\boldsymbol{\Pi}_m\hat{\mathbf{R}}^{(k)}_m\big\}\right)+\nonumber\\
&\frac{1}{\sigma_m}\mathrm{Tr}\left\{\big(\mathbf{I}_N-\boldsymbol{\Pi}_m\big)\hat{\mathbf{R}}^{(k)}_m\right\},\forall m.
\end{eqnarray}

Next, we simplify subproblems (28) by eliminating $\boldsymbol{\sigma}$. Thus, after estimating $\boldsymbol{\theta}$, $\boldsymbol{\sigma}$ is estimated by
\begin{eqnarray}       
\sigma^{(k)}_m=d^{(k)}_m/(N-1),\forall m.\nonumber
\end{eqnarray}
After eliminating $\boldsymbol{\sigma}$, subproblems (28) are simplified to
\begin{eqnarray}       
\min_{\theta_m\in\Omega}&(N-1)\ln\left(\mathrm{Tr}\big\{\hat{\mathbf{R}}^{(k)}_m\big\}-\mathrm{Tr}\big\{\boldsymbol{\Pi}_m\hat{\mathbf{R}}^{(k)}_m\big\}\right)\nonumber\\
&+\ln\left(\mathrm{Tr}\big\{\boldsymbol{\Pi}_m\hat{\mathbf{R}}^{(k)}_m\big\}\right),\forall m,
\end{eqnarray}
where $\theta_m\in\delta^{(k)}_m=\big\{\theta_m\in\Omega|\mathrm{Tr}\big\{\boldsymbol{\Pi}_m\hat{\mathbf{R}}^{(k)}_m\big\}>\frac{1}{N}\mathrm{Tr}\big\{\hat{\mathbf{R}}^{(k)}_m\big\}\big\}$ due to the fact that when $\theta^{(k)}_m\in\delta^{(k)}_m$, $e^{(k)}_m>\frac{1}{N}\mathrm{Tr}\big\{\hat{\mathbf{R}}^{(k)}_m\big\}$ and
\begin{eqnarray}       
r^{(k)}_m&=&\max\left\{\frac{1}{N}\big(e^{(k)}_m/\sigma^{(k)}_m-1\big),0\right\}\nonumber\\
&=&\max\left\{\frac{1}{N}\big[(N-1)e^{(k)}_m/d^{(k)}_m-1\big],0\right\}\nonumber\\
&=&\big(e^{(k)}_m-\frac{1}{N}\mathrm{Tr}\big\{\hat{\mathbf{R}}^{(k)}_m\big\}\big)/d^{(k)}_m>0,\forall m.\nonumber
\end{eqnarray}
Note that $(N-1)\ln\big(\mathrm{Tr}\big\{\hat{\mathbf{R}}^{(k)}_m\big\}-x\big)+\ln(x)$ is a monotonically decreasing function of $x$ for $x\ge\frac{1}{N}\mathrm{Tr}\{\hat{\mathbf{R}}^{(k)}_m\}$, subproblems (29) are thus equivalent to
\begin{eqnarray}       
\max_{\theta_m\in\delta^{(k)}_m}\mathrm{Tr}\big\{\boldsymbol{\Pi}_m\hat{\mathbf{R}}^{(k)}_m\big\},\forall m.
\end{eqnarray}

Based on the above analysis, the estimates of $\boldsymbol{\Theta}$ and $\boldsymbol{\sigma}$ are updated by
\begin{subequations}
\begin{eqnarray}     
\theta^{(k)}_m=
\arg\max_{\theta_m\in\Omega}\mathrm{Tr}\big\{\boldsymbol{\Pi}_m\hat{\mathbf{R}}^{(k)}_m\big\}, \forall m,
\end{eqnarray}
\begin{eqnarray}       
\sigma^{(k)}_m=
\left\{
\begin{array}{ll}
{d^{(k)}_m/(N-1)} & {e^{(k)}_m>\frac{1}{N}\mathrm{Tr}\big\{\hat{\mathbf{R}}^{(k)}_m\big\}}, \\
{\frac{1}{N}\mathrm{Tr}\big\{\hat{\mathbf{R}}^{(k)}_m\big\}} & {e^{(k)}_m\le \frac{1}{N}\mathrm{Tr}\big\{\hat{\mathbf{R}}^{(k)}_m\big\}}, \\
\end{array}
\right.
\forall m,
\end{eqnarray}
\begin{eqnarray}       
P^{(k)}_m=\max\left\{\frac{1}{N-1}\big(e^{(k)}_m-\frac{1}{N}\mathrm{Tr}\big\{\hat{\mathbf{R}}^{(k)}_m\big\}\big),0\right\},\forall m.
\end{eqnarray}
\end{subequations}

Finally, we give the following remark.

\begin{remark}
In the MEM algorithm for the stochastic signal model, $\boldsymbol{\sigma}^{(k)}>\mathbf{0}$ for $k\ge1$ if $\boldsymbol{\sigma}^{(0)}>\mathbf{0}$.

\end{remark}

\begin{proof}
We utilize a proof by contradiction. Without loss of generality, assume $\boldsymbol{\sigma}^{(k)}>\mathbf{0}$ for $0\le k\le K-1$ $(K\ge1)$ and $\sigma^{(K)}_i=0$. Obviously, we have
\begin{eqnarray}       
\mathbf{C}^{(K-1)}_m=&P^{(K-1)}_m\mathbf{a}(\theta^{(K-1)}_m)\mathbf{a}^H(\theta^{(K-1)}_m)\nonumber\\
&+\sigma^{(K-1)}_m\mathbf{I}_N>\mathbf{0}_N,\forall m.
\end{eqnarray}

Based on (31b) and $\sigma^{(K)}_i=0$, we first consider $\sigma^{(K)}_i=\frac{1}{N}\mathrm{Tr}\big\{\hat{\mathbf{R}}^{(K)}_i\big\}=0$. Then, $\hat{\mathbf{R}}^{(K)}_i=\mathbf{0}_N$ as $\hat{\mathbf{R}}^{(K)}_i\ge\mathbf{0}_N$, resulting to that $e^{(K)}_i=0$ and by (25) we have
\begin{eqnarray}       
\mathbf{C}^{(K-1)}_i[\mathbf{C}^{(K-1)}_y]^{-1}\hat{\mathbf{R}}_y[\mathbf{C}^{(K-1)}_y]^{-1}\mathbf{C}^{(K-1)}_i&=&\mathbf{0}_N,\nonumber\\ \mathbf{C}^{(K-1)}_i-\mathbf{C}^{(K-1)}_i[\mathbf{C}^{(K-1)}_y]^{-1}\mathbf{C}^{(K-1)}_i&=&\mathbf{0}_N,\nonumber
\end{eqnarray}
which indicate $\hat{\mathbf{R}}_y=\mathbf{0}_N$ and $\mathbf{C}^{(K-1)}_i=\mathbf{C}^{(K-1)}_y=\sum_{m=1}^M\mathbf{C}^{(K-1)}_m$. Thus, $\sum_{m\ne i}\mathbf{C}^{(K-1)}_m=\mathbf{0}_N$, which contradicts $\sum_{m\ne i}\mathbf{C}^{(K-1)}_m>\mathbf{0}_N$ by (32).

Next, we consider $\sigma^{(K)}_i=d^{(K)}_i/(N-1)=0$, i.e., $\hat{\mathbf{R}}^{(K)}_i=f^{(K-1)}_i\boldsymbol{\Pi}_i>\mathbf{0}_N$ as $e^{(K)}_i>\frac{1}{N}\mathrm{Tr}\big\{\hat{\mathbf{R}}^{(K)}_i\big\}$. Then, by (25) we have
\begin{eqnarray}       
\mathbf{C}^{(K-1)}_i[\mathbf{C}^{(K-1)}_y]^{-1}\hat{\mathbf{R}}_y[\mathbf{C}^{(K-1)}_y]^{-1}\mathbf{C}^{(K-1)}_i&=&
h^{(K-1)}_i\boldsymbol{\Pi}_i,\nonumber\\
\mathbf{C}^{(K-1)}_i-\mathbf{C}^{(K-1)}_i[\mathbf{C}^{(K-1)}_y]^{-1}\mathbf{C}^{(K-1)}_i&=&
v^{(K-1)}_i\boldsymbol{\Pi}_i,\nonumber
\end{eqnarray}
where $h^{(K-1)}_i\ge0$ and $v^{(K-1)}_i>0$, resulting in
\begin{eqnarray}       
\left\vert\mathbf{C}^{(K-1)}_i-\mathbf{C}^{(K-1)}_i[\mathbf{C}^{(K-1)}_y]^{-1}\mathbf{C}^{(K-1)}_i\right\vert=0,\nonumber
\end{eqnarray}
and
\begin{eqnarray}       
\left\vert\mathbf{C}^{(K-1)}_y-\mathbf{C}^{(K-1)}_i\right\vert=\left\vert{\sum}_{m\ne i}\mathbf{C}^{(K-1)}_m\right\vert=0,\nonumber
\end{eqnarray}
which contradicts $\sum_{m\ne i}\mathbf{C}^{(K-1)}_m>\mathbf{0}_N$ by (32). The proof is completed.
\end{proof}

\section{SAGE Algorithm}

In this section, the SAGE algorithm is proposed. 
We design that the SAGE algorithm updates the DOA estimates in the ascending order of source number at each iteration and one iteration finishes when all the parameter estimates are updated. Besides, let $(\cdot)^{(k,i)}$ denote an iterative value obtained during updating the estimate of $\theta_i$ at the $k$th iteration, $(\cdot)^{(k-1)}=(\cdot)^{(k-1,M)}=(\cdot)^{(k,0)}$.

The SAGE algorithm updates the parameter estimates by the E- and M-steps of the EM or MEM algorithm. When updating the estimate of $\theta_i$ at the $k$th iteration, the SAGE algorithm first associates all of the noise with the $i$th source signal component \cite{bibitem6}, \cite{bibitem7}:
\begin{equation}          
\alpha_m=
\left\{
\begin{array}{ll}
{1} & {m=i}, \\
{0} & {m\ne i}. \\
\end{array}
\right.
\end{equation}

\subsection{Deterministic Signal Model}
According to (33), $\mathbf{x}_i(t)\sim\mathcal{CN}(\mathbf{a}(\theta_i)s_i(t),\sigma\mathbf{I}_N)$, $\mathbf{x}_m(t)=\mathbf{a}(\theta_m)s_m(t)$ for $m\ne i$, and (3b) is rewritten as
\begin{eqnarray}       
&\mathcal{Q}(\mathbf{X}_i,\theta_i,\mathbf{s}_i,\sigma)=\ln p(\mathbf{X}_1,\dots,\mathbf{X}_M;\boldsymbol{\theta},\mathbf{S},\sigma)\nonumber\\
=&-TN\ln(\pi\sigma)-\frac{1}{\sigma}\sum_{t=1}^T\Vert\mathbf{x}_i(t)-\mathbf{a}(\theta_i)s_i(t)\Vert^2.
\end{eqnarray}

At the E-step, the SAGE algorithm computes the conditional expectation of $\mathcal{Q}(\mathbf{X}_i,\theta_i,\mathbf{s}_i,\sigma)$ by
\begin{eqnarray}       
&\mathbb{E}\left\{\mathcal{Q}(\mathbf{X}_i,\theta_i,\mathbf{s}_i,\sigma)|\mathbf{Y};\boldsymbol{\Theta}^{(k,i-1)},\sigma^{(k,i-1)}\right\}\nonumber\\
=&-TN\ln(\pi\sigma)-\frac{1}{\sigma}\sum_{t=1}^T\Vert\mathbf{x}^{(k)}_i(t)-\mathbf{a}(\theta_i)s_i(t)\Vert^2,
\end{eqnarray}
where $\boldsymbol{\Theta}=(\boldsymbol{\theta},\mathbf{S})$ and
\begin{subequations}
\begin{eqnarray}       
\mathbf{x}_i^{(k)}(t)=&\mathbf{x}_i^{(k,i)}(t)=\mathbb{E}\left\{\mathbf{x}_i(t)|\mathbf{Y};\boldsymbol{\Theta}^{(k,i-1)},\sigma^{(k,i-1)}\right\}\nonumber\\
=&\big[\mathbf{y}(t)-\mathbf{A}(\boldsymbol{\theta}^{(k,i-1)})\mathbf{s}^{(k,i-1)}(t)\big]\nonumber\\
&+\mathbf{a}(\theta_i^{(k,i-1)})s_i^{(k,i-1)}(t),\forall t,
\end{eqnarray}
\begin{eqnarray}       
\mathbb{D}\left\{\mathbf{x}_i(t)|\mathbf{Y};\boldsymbol{\Theta}^{(k,i-1)},\sigma^{(k,i-1)}\right\}=\mathbf{0}_N,\forall t.
\end{eqnarray}
\end{subequations}

At the M-step, based on (35) the SAGE algorithm estimates $\theta_i$, $\mathbf{s}_i$, and $\sigma$ by
\begin{eqnarray}       
\min_{\theta_i\in\Omega,\mathbf{s}_i,\sigma>0}N\ln(\sigma)+\frac{1}{\sigma T}\sum_{t=1}^T\Vert\mathbf{x}_i^{(k)}(t)-\mathbf{a}(\theta_i)s_i(t)\Vert^2.
\end{eqnarray}
Thus, the estimates of $\theta_i$, $\mathbf{s}_i$, and $\sigma$ are updated by
\begin{subequations}
\begin{eqnarray}       
\theta^{(k)}_i=\theta^{(k,i)}_i=\arg\max_{\theta_i\in\Omega}\mathrm{Tr}\big\{\boldsymbol{\Pi}_i\hat{\mathbf{R}}^{(k)}_i\big\},
\end{eqnarray}
\begin{eqnarray}       
s^{(k)}_i(t)=s^{(k,i)}_i(t)=\frac{1}{N}\mathbf{a}^H(\theta^{(k)}_i)\mathbf{x}^{(k)}_i(t),\forall t,
\end{eqnarray}
\begin{eqnarray}       
\sigma^{(k,i)}=d^{(k)}_i/N,
\end{eqnarray}
\end{subequations}
where $\hat{\mathbf{R}}^{(k)}_i=\frac{1}{T}\sum_{t=1}^T\mathbf{x}^{(k)}_i(t)[\mathbf{x}^{(k)}_i(t)]^H$.

Finally, the other parameter estimates are not updated and their iterative values are
\begin{subequations}
\begin{eqnarray}       
\theta^{(k,i)}_m=\theta^{(k,i-1)}_m,\forall m\ne i,
\end{eqnarray}
\begin{eqnarray}       
s^{(k,i)}_m(t)=s^{(k,i-1)}_m(t),\forall m\ne i, t.
\end{eqnarray}
\end{subequations}

At each iteration of the SAGE algorithm, the E- and M-steps are repeated $M$ times, leading to that $\mathbf{X}_1,\dots,\mathbf{X}_M$ and all elements in $\boldsymbol{\Theta}$ are estimated once while $\sigma$ is estimated $M$ times.

\begin{remark}
Notice that the $\mathbf{x}^{(k)}_i(t)$'s in (36a), $\theta^{(k)}_i$ in (38a), and the $s^{(k)}_i(t)$'s in (38b) are unrelated to $\sigma^{(k,i-1)}$ and respectively identical with these in the SAGE algorithm under the known $\sigma$ \cite{bibitem8}, i.e., the SAGE algorithm under the unknown $\sigma$ is equivalent to that under the known $\sigma$ when not estimating $\sigma$. Accordingly, (38c) can be omitted when the algorithm does not consider the nuisance parameter $\sigma$.
\end{remark}


\subsection{Stochastic Signal Model}
According to (33), $\mathbf{x}_i(t)\sim\mathcal{CN}(\textbf{0},\mathbf{C}_i)$ with $\mathbf{C}_i=P_i\mathbf{a}(\theta_i)\mathbf{a}^H(\theta_i)+\sigma\mathbf{I}_N$ and for $m\ne i$, the statistical model of $\mathbf{x}_m(t)$ depends only on $s_m(t)$ due to $\mathbf{x}_m(t)=\mathbf{a}(\theta_m)s_m(t)$. Thus, (5b) is rewritten as
\begin{eqnarray}       
&\mathcal{Q}(\mathbf{S}_i,\mathbf{X}_i,\theta_i,\mathbf{P},\sigma)=\ln p(\mathbf{X}_1,\dots,\mathbf{X}_M;\boldsymbol{\theta},\mathbf{P},\sigma)\nonumber \\
=&-T(M-1)\ln(\pi)-T\sum_{m\ne i}\big[\ln(P_m)+\hat{P}_m/P_m\big]\nonumber \\
&-TN\ln(\pi)-T\big[\ln\vert\mathbf{C}_i\vert+\mathrm{Tr}\big\{\mathbf{C}_i^{-1}\hat{\mathbf{R}}_i\big\}\big],
\end{eqnarray}
where $\vert a\vert$ denotes the modulus of a complex number $a$, $\hat{P}_m$ is a sufficient statistic for $P_m$ \cite{bibitem16},
\begin{eqnarray}
\mathbf{S}_i&=&[\cdots~\mathbf{s}_{i-2}^T~\mathbf{s}_{i-1}^T~\mathbf{s}_{i+1}^T~\mathbf{s}_{i+2}^T~\cdots]^T,\nonumber \\
\hat{P}_m&=&\frac{1}{T}{\sum}_{t=1}^{T}\vert s_m(t)\vert^2.\nonumber
\end{eqnarray}

At the E-step, the SAGE algorithm computes the conditional expectation of $\mathcal{Q}(\mathbf{S}_i,\mathbf{X}_i,\theta_i,\mathbf{P},\sigma)$ by
\begin{eqnarray}       
&\mathbb{E}\left\{\mathcal{Q}(\mathbf{S}_i,\mathbf{X}_i,\theta_i,\mathbf{P},\sigma)|\mathbf{Y};\boldsymbol{\Theta}^{(k,i-1)},\sigma^{(k,i-1)}\right\}\nonumber\\
=&-T(M-1)\ln(\pi)-TN\ln(\pi)\nonumber \\
&-T\sum_{m\ne i}\big[\ln(P_m)+\hat{P}^{(k,i)}_m/P_m\big]\nonumber \\
&-T\big[\ln\vert\mathbf{C}_i\vert+\mathrm{Tr}\big\{\mathbf{C}_i^{-1}\hat{\mathbf{R}}^{(k)}_i\big\}\big],
\end{eqnarray}
where $\boldsymbol{\Theta}=(\boldsymbol{\theta},\mathbf{P})$,
\begin{eqnarray}       
\hat{P}^{(k,i)}_m=&\mathbb{E}\left\{\hat{P}_m|\mathbf{Y};\boldsymbol{\Theta}^{(k,i-1)},\sigma^{(k,i-1)}\right\}\nonumber\\
=&P^{(k,i-1)}_m\big[1-\mathbf{a}^H(\theta^{(k,i-1)}_m)\mathbf{b}^{(k,i-1)}_m\big]\nonumber\\
&+[\mathbf{b}^{(k,i-1)}_m]^H\hat{\mathbf{R}}_y\mathbf{b}^{(k,i-1)}_m\ge0
\end{eqnarray}
with $\mathbf{b}^{(k,i-1)}_m=[\mathbf{C}^{(k,i-1)}_y]^{-1}\mathbf{a}(\theta^{(k,i-1)}_m)P^{(k,i-1)}_m$, and
\begin{eqnarray}       
\hat{\mathbf{R}}^{(k)}_i=&\hat{\mathbf{R}}^{(k,i)}_i=\mathbb{E}\left\{\hat{\mathbf{R}}_i|\mathbf{Y};\boldsymbol{\Theta}^{(k,i-1)},\sigma^{(k,i-1)}\right\}\nonumber\\
=&\mathbf{C}^{(k,i-1)}_i[\mathbf{C}^{(k,i-1)}_y]^{-1}\hat{\mathbf{R}}_y[\mathbf{C}^{(k,i-1)}_y]^{-1}\mathbf{C}^{(k,i-1)}_i+\nonumber\\
&\left\{\mathbf{C}^{(k,i-1)}_i-\mathbf{C}^{(k,i-1)}_i[\mathbf{C}^{(k,i-1)}_y]^{-1}\mathbf{C}^{(k,i-1)}_i\right\}.
\end{eqnarray}

At the M-step, based on (41) the SAGE algorithm updates the estimates of $\theta_i$, $\mathbf{P}$, and $\sigma$ by
\begin{eqnarray}       
\min_{\theta_i\in\Omega,\mathbf{P}\ge\mathbf{0},\sigma>0}&\sum_{m\ne i}\big[\ln(P_m)+\hat{P}^{(k,i)}_m/P_m\big]\nonumber \\
&+\big[\ln\vert\mathbf{C}_i\vert+\mathrm{Tr}\big\{\mathbf{C}_i^{-1}\hat{\mathbf{R}}^{(k)}_i\big\}\big],
\end{eqnarray}
resulting in that
\begin{eqnarray}       
P^{(k,i)}_m=\hat{P}^{(k,i)}_m,\forall m\ne i,
\end{eqnarray}
while the estimates of $\theta_i$, $P_i$, and $\sigma$ are updated by
\begin{eqnarray}       
\min_{\theta_i\in\Omega,r_i\ge0,\sigma>0}N\ln(\sigma)+\ln\vert\mathbf{D}_i\vert+\frac{1}{\sigma}\mathrm{Tr}\big\{\mathbf{D}_i^{-1}\hat{\mathbf{R}}^{(k)}_i\big\},
\end{eqnarray}
where $\mathbf{C}_i=\sigma\mathbf{D}_i$ with $\mathbf{D}_i=r_i\mathbf{a}(\theta_i)\mathbf{a}^H(\theta_i)+\mathbf{I}_N$ and the solution can be obtained from the solutions of subproblems (26). Following (31), $\theta_i$, $P_i$, and $\sigma$ can be estimated by
\begin{subequations}
\begin{eqnarray}     
\theta^{(k)}_i=\theta^{(k,i)}_i=
\arg\max_{\theta_i\in\Omega}\mathrm{Tr}\big\{\boldsymbol{\Pi}_i\hat{\mathbf{R}}^{(k)}_i\big\},
\end{eqnarray}
\begin{eqnarray}       
\sigma^{(k,i)}=
\left\{
\begin{array}{ll}
{d^{(k)}_i/(N-1)} & {e^{(k)}_i>\frac{1}{N}\mathrm{Tr}\big\{\hat{\mathbf{R}}^{(k)}_i\big\}}, \\
{\frac{1}{N}\mathrm{Tr}\big\{\hat{\mathbf{R}}^{(k)}_i\big\}} & {e^{(k)}_i\le\frac{1}{N}\mathrm{Tr}\big\{\hat{\mathbf{R}}^{(k)}_i\big\}}, \\
\end{array}
\right.
\end{eqnarray}
\begin{eqnarray}       
P^{(k,i)}_i=\max\left\{\frac{1}{N-1}\big(e^{(k)}_i-\frac{1}{N}\mathrm{Tr}\big\{\hat{\mathbf{R}}^{(k)}_i\big\}\big),0\right\},
\end{eqnarray}
\end{subequations}
where $\sigma^{(k,i)}=0$ is possible although its probability is very low. For example, if $M=2$, $T=1$ (a single snapshot), and $P^{(1,1)}_1=0$ after updating the estimate of $\theta_1$, we will have
\begin{eqnarray}
\mathbf{C}^{(1,1)}_2=\mathbf{C}^{(1,1)}_y=P^{(1,1)}_2\mathbf{a}(\theta^{(1,1)}_2)\mathbf{a}^H(\theta^{(1,1)}_2)+\sigma^{(1,1)}\mathbf{I}_N>\mathbf{0}_N \nonumber
\end{eqnarray}
and $\hat{\mathbf{R}}^{(1)}_2=\mathbf{y}(1)\mathbf{y}^H(1)$ by (43) when updating the estimate of $\theta_2$. Furthermore, if $\mathbf{y}(1)=u\mathbf{a}(\bar{\theta})$ $(u\ne0)$, we will obtain $\theta^{(1)}_2=\bar{\theta}$ by (47a) and $\sigma^{(1,2)}=d^{(1)}_2/(N-1)=0$ by (47b).

To avoid $\sigma^{(k,i)}=0$, we use two CM-steps based on the ECM algorithm to reestimate $\theta_i$, $P_i$, and $\sigma$ by problem (46) if $\sigma^{(k,i)}=0$ in (47b).
\begin{itemize}
\item \emph{First CM-step:} estimate $\theta_i$ and $P_i$ but hold $\sigma=\sigma^{(k,i-1)}$ fixed. Then, problem (46) is simplified to
\begin{eqnarray} 
\min_{\theta_i\in\Omega,r_i\ge0}\ln\vert\mathbf{D}_i\vert+\frac{1}{\sigma^{(k,i-1)}}\mathrm{Tr}\big\{\mathbf{D}_i^{-1}\hat{\mathbf{R}}^{(k)}_i\big\},
\end{eqnarray}
which can be solved by referring to (15). Thus, the estimates of $\theta_i$ and $P_i$ are updated by
\begin{subequations}
\begin{eqnarray}     
\theta^{(k)}_i=\theta^{(k,i)}_i=
\arg\max_{\theta_i\in\Omega}\mathrm{Tr}\big\{\boldsymbol{\Pi}_i\hat{\mathbf{R}}^{(k)}_i\big\},
\end{eqnarray}
\begin{eqnarray}       
P^{(k,i)}_i=&r^{(k,i)}_i\sigma^{(k,i-1)}\nonumber\\
=&\max\left\{\frac{1}{N}\big(e^{(k)}_i-\sigma^{(k,i-1)}\big),0\right\},
\end{eqnarray}
\end{subequations}
where $\theta^{(k)}_i$ is indeterminate if $P^{(k,i)}_i=0$.
\item \emph{Second CM-step:} estimate $\sigma$ but hold $\theta_i=\theta^{(k)}_i$ and $r_i=r^{(k,i)}_i$ fixed. Then, problem (46) is simplified to
\begin{eqnarray} 
\min_{\sigma>0}N\ln(\sigma)+\frac{1}{\sigma}\mathrm{Tr}\left\{[\mathbf{D}^{(k)}_i]^{-1}\hat{\mathbf{R}}^{(k)}_i\right\},
\end{eqnarray}
where $\mathbf{D}^{(k)}_i=r^{(k,i)}_i\mathbf{a}(\theta^{(k)}_i)\mathbf{a}^H(\theta^{(k)}_i)+\mathbf{I}_N$. Thus, the estimate of $\sigma$ is updated by
\begin{eqnarray} 
\sigma^{(k,i)}&=&\frac{1}{N}\mathrm{Tr}\left\{[\mathbf{D}^{(k)}_i]^{-1}\hat{\mathbf{R}}^{(k)}_i\right\}\nonumber\\
&=&\big(\sigma^{(k,i-1)}+d^{(k)}_i\big)/N,
\end{eqnarray}
which implies $\sigma^{(k,i)}>0$ if $\sigma^{(k,i-1)}>0$.

\end{itemize}

Finally, the other parameter estimate(s) is(are) not updated and the iterative value(s) is(are)
\begin{eqnarray}     
\theta^{(k,i)}_m=\theta^{(k,i-1)}_m,\forall m\ne i.
\end{eqnarray}

The E- and M-steps are repeated $M$ times at each iteration of the SAGE algorithm, so $\hat{\mathbf{R}}_1,\dots,\hat{\mathbf{R}}_M$ and all elements in $\boldsymbol{\theta}$ are estimated once, $\sigma$ and all elements in $\mathbf{P}$ are estimated $M$ times, each element of $\hat{\mathbf{P}}=[\hat{P}_1~\cdots~\hat{P}_M]^T$ is estimated $M-1$ times or once ($M=2$).

\section{Properties of the Proposed EM, MEM, and SAGE Algorithms}

\subsection{Convergence Point}

It is well known that under the known $\sigma$,
the EM and SAGE algorithms satisfy standard regularity conditions \cite{bibitem4}, \cite{bibitem6}, \cite{bibitem20} and converge to different stationary points or the same stationary point of $\mathcal{L}(\boldsymbol{\Theta})$, which is $\mathcal{L}(\boldsymbol{\Theta},\sigma)$ with $\sigma$ fixed, implying that $\mathcal{L}(\boldsymbol{\Theta},\sigma)$ is a well-behaved objective function. Thus, the proposed algorithms also satisfy the regularity conditions and converge to different stationary points or the same stationary point of $\mathcal{L}(\boldsymbol{\Theta},\sigma)$.

Of course, the convergence points of the proposed algorithms depend on their initial points.
Given a poor initial point, the proposed algorithms may never converge to the maximum point of $\mathcal{L}(\boldsymbol{\Theta},\sigma)$. To generate an appropriate initial point, the effective initialization procedure in \cite{bibitem9} can be adopted using the deterministic signal model.

\subsection{Complexity and Stability}

Note that the computational complexities of the EM, MEM, and SAGE algorithms are dominated by searching the $\theta_m^{(k)}$'s:
\begin{eqnarray} 
\theta_m^{(k)}=\arg\max_{\theta_m\in\Omega}\mathrm{Tr}\big\{\boldsymbol{\Pi}_m\hat{\mathbf{R}}_m^{(k)}\big\},\forall m.
\end{eqnarray}
Hence, if we adopt brute force to search $\theta_m^{(k)}$, i.e., evaluating the objective function $\mathrm{Tr}\{\boldsymbol{\Pi}_m\hat{\mathbf{R}}_m^{(k)}\}$ on a coarse grid to locate a grid point, close to the maximum point of $\mathrm{Tr}\{\boldsymbol{\Pi}_m\hat{\mathbf{R}}_m^{(k)}\}$, as the initial point of a gradient algorithm and then applying this gradient algorithm to search the maximum point as $\theta_m^{(k)}$ \cite{bibitem4}, these algorithms will have almost the same computational complexity at each iteration \cite{bibitem7}.

However, when the powers of sources are unequal, we have found via simulation that the DOA estimates of multiple sources, updated by (53), tend to be consistent with the true DOA of the source with the largest power and the algorithms may be unstable.

To address this issue, we can choose $\theta_m^{(k-1)}$ as the initial point of a gradient algorithm and then apply this gradient algorithm to search a local maximum point of $\mathrm{Tr}\big\{\boldsymbol{\Pi}_m\hat{\mathbf{R}}_m^{(k)}\big\}$ as $\theta_m^{(k)}$, e.g., \textbf{Algorithm 2} in the next section has given excellent numerical results. Under this choice, we still have
\begin{eqnarray} 
\mathrm{Tr}\big\{\boldsymbol{\Pi}^{(k)}_m\hat{\mathbf{R}}_m^{(k)}\big\}\ge
\mathrm{Tr}\big\{\boldsymbol{\Pi}^{(k-1)}_m\hat{\mathbf{R}}_m^{(k)}\big\},\forall m,
\end{eqnarray}
which actually guarantee the monotonicity of the algorithms but only meet the requirement of GEM algorithms \cite{bibitem3}. For convenience, we do not change the names of the algorithms under this choice.

\section{Numerical Results}

In this section, numerical results are provided to illustrate the convergence performances of the proposed algorithms. To ensure that a two-dimensional scatter plot can reflect the DOAs of the $M$ sources, the array is assumed to be a uniform linear array with $\mathbf{p}_{n} = [\frac{\lambda}{2}(n-1)~0~0]^{T}$, $N=10$, and $T=20$, then $M=2$ and $\varphi_1=\varphi_2=90\degree$ are known while $\phi_1$ and $\phi_2$ are to be estimated. SAGE\footnote{In this section, the EM, MEM, and SAGE algorithms are simply written as EM, MEM, and SAGE, respectively.} for the stochastic signal model is given in \textbf{Algorithm 1} and the other algorithms in this section can be obtained by referring to \textbf{Algorithm 1}. For comparison, let all tolerances be $\epsilon=0.001$, and $\mathbf{S}$ in the deterministic signal model is also generated by the independent random numbers $s_m(t)\sim\mathcal{CN}(0,P_m)$.

\begin{algorithm}
\caption{SAGE for the Stochastic Signal Model} \label{alg:A}
\begin{algorithmic}[1]
\STATE {Initialize the parameter set $\xi^{(0)}=\{\boldsymbol{\theta}^{(0)},\mathbf{P}^{(0)},\sigma^{(0)}\}$ with $\boldsymbol{\theta}=[\phi_1~\phi_2]^T$, $k=1$, and $i=1$.}
\WHILE {$i\le M$}
       \STATE {Find $\hat{\mathbf{R}}^{(k)}_i$ and $\hat{P}^{(k,i)}_m$ for $m\ne i$ at the E-step and obtain $\xi^{(k,i)}$ at the M-step, then $i=i+1$.}
\ENDWHILE
\WHILE 
{$\Vert\boldsymbol{\theta}^{(k)}-\boldsymbol{\theta}^{(k-1)}\Vert>\epsilon~(\mathrm{degree})$}
       \STATE {$k=k+1$ and $i=1$.}
       \WHILE {$i\le M$}
       \STATE {Find $\hat{\mathbf{R}}^{(k)}_i$ and $\hat{P}^{(k,i)}_m$ for $m\ne i$ at the E-step and obtain $\xi^{(k,i)}$ at the M-step, then $i=i+1$.}
       \ENDWHILE
\ENDWHILE
\STATE {Output $\boldsymbol{\theta}^{(k)}$.}
\end{algorithmic}
\end{algorithm}

Furthermore, $\boldsymbol{\alpha}=[0.5~0.5]^T$ in the EM algorithm and the gradient ascent method with the backtracking line search \cite{bibitem21}, given in \textbf{Algorithm 2}, is adopted to search the $\phi^{(k)}_m$'s in (54).
Several simulation parameters in \textbf{Algorithm 2} are $\rho=0.1$, $\eta=0.3$, and $\gamma=0.5$.

\begin{algorithm}
\caption{Gradient Ascent Based Angle Estimation} \label{alg:A}
\begin{algorithmic}[1]
\STATE {Define $g(\phi_m)=N\times\mathrm{Tr}\big\{\boldsymbol{\Pi}_m\hat{\mathbf{R}}^{(k)}_m\big\}$, initialize $\phi_m=\phi_m^{(k-1)}\in(0,\pi)~(\mathrm{radian})$, $\rho\in(0,1)$, $\eta\in(0,0.5)$, and $\gamma\in(0,1)$.}
        \WHILE{$\vert g'(\phi_m)\vert>\epsilon$}
           \STATE {$t=\left\{
\begin{array}{ll}
\rho\times\frac{\pi-\phi_m}{g'(\phi_m)},&g'(\phi_m)>0,\\
\rho\times\frac{-\phi_m}{g'(\phi_m)},&g'(\phi_m)<0.
\end{array}
\right.$}
           \WHILE{$g\big(\phi_m+tg'(\phi_m)\big)<g(\phi_m)+\eta t\vert g'(\phi_m)\vert^2$}
           \STATE {$t=\gamma t$.}
           \ENDWHILE
           \STATE {$\phi_m=\phi_m+tg'(\phi_m)$.}
       \ENDWHILE
\STATE {$\phi_m^{(k)}=\phi_m$.}
\end{algorithmic}
\end{algorithm}

\begin{subsection}{Deterministic Signal Model}

For comparing the convergence of EM, MEM, and SAGE, Fig. 1 plots their $\mathcal{L}(\boldsymbol{\Theta}^{(k)},\sigma^{(k)})$'s, $\phi^{(k)}_1$'s, and $\phi^{(k)}_2$'s as functions of $k$ under one realization.
We observe that given a good initial point, 
the algorithms can converge to consistent DOA estimates,
and EM has similar convergence with MEM while SAGE converges faster than EM and MEM.

\begin{figure}[t] \centering
\includegraphics[scale=0.6]{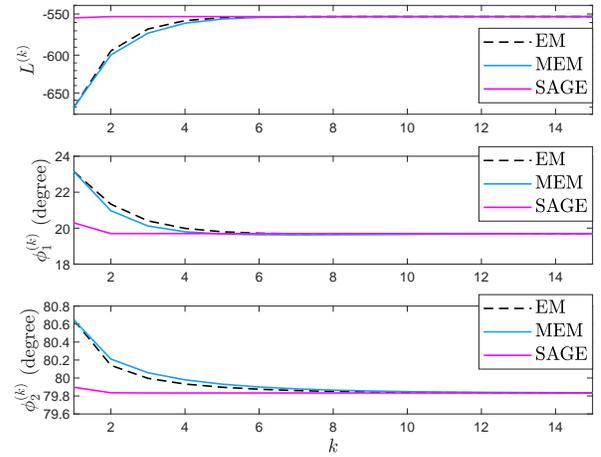}
\vspace{0cm}\caption{$\mathcal{L}(\boldsymbol{\Theta}^{(k)},\sigma^{(k)})$, $\phi^{(k)}_1$, and $\phi^{(k)}_2$ comparison of EM, MEM, and SAGE for the deterministic signal model under one realization versus $k$ with $\phi_1=20\degree$, $\phi_2=80\degree$, $P_1=-2~\mathrm{dB}$, $P_2=4~\mathrm{dB}$, $\sigma=4~\mathrm{dB}$, $\phi^{(0)}_1=24\degree$, $\phi^{(0)}_2=84\degree$, $\mathbf{S}^{(0)}=[\mathbf{1}~\mathbf{1}]^T$, $\boldsymbol{\sigma}^{(0)}=[0.5~0.5]^T$, and $\sigma^{(0)}=1$.}\vspace{0cm}
\end{figure}


Figs. 2 and 3 show two scatter plots of the DOA estimates obtained by the algorithms under 200 independent realizations. The same samples of each realization are processed by the algorithms. In Fig. 2, the total numbers of wanted points from EM, MEM, and SAGE are 68, 72, and 179, respectively. In Fig. 3, the total numbers of wanted points from EM, MEM, and SAGE are 159, 157, and 190, respectively. Figs. 2 and 3 imply that given a poor initial point, SAGE is more efficient for avoiding the convergence to an unwanted stationary point of $\mathcal{L}(\boldsymbol{\Theta},\sigma)$ than EM and MEM.


Note that both sources in Fig. 2 are not closely spaced, so it is very difficult to mix up both sources and the wanted points in Fig. 2 are centered around the true position $(25\degree,75\degree)$. However, both sources in Fig. 3 are closely spaced and the wanted points are centered around $(78\degree,70\degree)$ or $(70\degree,78\degree)$,
i.e., the algorithms are likely to mix up closely spaced sources. Fortunately, we only focus on the DOAs of sources in DOA estimation and mixing up sources has no effect on the result.

\begin{figure}[t] \centering
\includegraphics[scale=0.6]{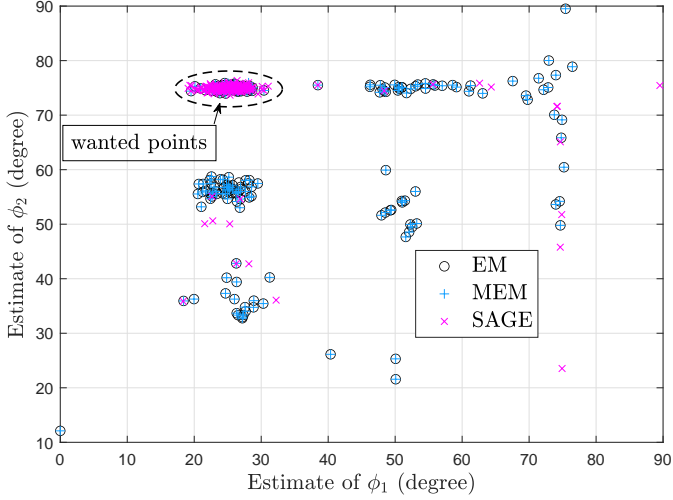}
\vspace{0cm}\caption{Scatter plot of the DOA estimates obtained from EM, MEM, and SAGE for the deterministic signal model under 200 independent realizations with $\phi_1=25\degree$, $\phi_2=75\degree$, $P_1=-4~\mathrm{dB}$, $P_2=2~\mathrm{dB}$, $\sigma=4~\mathrm{dB}$, $\phi^{(0)}_1=40\degree$, $\phi^{(0)}_2=60\degree$, $\mathbf{S}^{(0)}=[\mathbf{1}~\mathbf{1}]^T$, $\boldsymbol{\sigma}^{(0)}=[0.5~0.5]^T$.}\vspace{0cm}
\end{figure}

\begin{figure}[t] \centering
\includegraphics[scale=0.6]{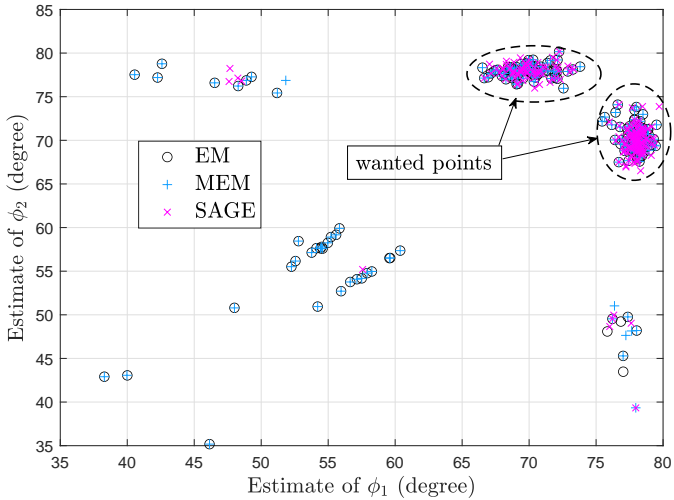}
\vspace{0cm}\caption{Scatter plot of the DOA estimates obtained from EM, MEM, and SAGE for the deterministic signal model under 200 independent realizations with $\phi_1=70\degree$, $\phi_2=78\degree$, $P_1=-2~\mathrm{dB}$, $P_2=4~\mathrm{dB}$, $\sigma=4~\mathrm{dB}$, $\phi^{(0)}_1=50\degree$, $\phi^{(0)}_2=58\degree$, $\mathbf{S}^{(0)}=[\mathbf{1}~\mathbf{1}]^T$, $\boldsymbol{\sigma}^{(0)}=[0.5~0.5]^T$.}\vspace{0cm}
\end{figure}

According to these simulations, we can conclude that for the deterministic signal model, 1) EM has similar convergence with MEM, and 2) SAGE outperforms EM and MEM.

\end{subsection}

\begin{subsection}{Stochastic Signal Model}

For comparing the convergence of EM, MEM, and SAGE, Fig. 4 plots their $\mathcal{L}(\boldsymbol{\Theta}^{(k)},\sigma^{(k)})$'s, $\phi^{(k)}_1$'s, and $\phi^{(k)}_2$'s as functions of $k$. We can observe that EM has similar convergence with MEM while SAGE converges faster than EM and MEM.

\begin{figure}[t] \centering
\includegraphics[scale=0.6]{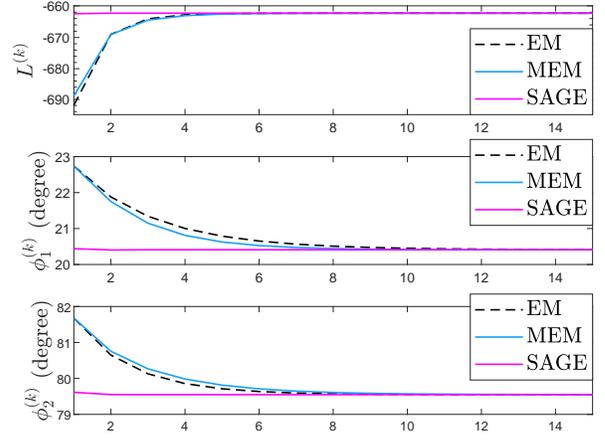}
\vspace{0cm}\caption{$\mathcal{L}(\boldsymbol{\Theta}^{(k)},\sigma^{(k)})$, $\phi^{(k)}_1$, and $\phi^{(k)}_2$ comparison of EM, MEM, and SAGE for the stochastic signal model under one realization versus $k$ with $\phi_1=20\degree$, $\phi_2=80\degree$, $P_1=-4~\mathrm{dB}$, $P_2=4~\mathrm{dB}$, $\sigma=4~\mathrm{dB}$, $\phi^{(0)}_1=24\degree$, $\phi^{(0)}_2=84\degree$, $\mathbf{P}^{(0)}=\mathbf{1}$, $\boldsymbol{\sigma}^{(0)}=[0.5~0.5]^T$, and $\sigma^{(0)}=1$.}\vspace{0cm}
\end{figure}


Figs. 5 and 6 show two scatter plots of the DOA estimates obtained by the algorithms under 200 independent realizations. The same samples of each realization are processed by the algorithms. In Fig. 5, the total numbers of wanted points from EM, MEM, and SAGE are 185, 186, and 175, respectively. In Fig. 6, the total numbers of wanted points from EM, MEM, and SAGE are 161, 161, and 172, respectively. Figs. 5 and 6 indicate that EM has similar convergence with MEM but compared to EM and MEM, SAGE is less and more efficient for avoiding the convergence to an unwanted stationary point of $\mathcal{L}(\boldsymbol{\Theta},\sigma)$ in Figs. 5 and 6, respectively.
The algorithms are likely to mix up closely spaced sources, so the wanted points in Fig. 5 are centered around $(25\degree,75\degree)$ and the wanted points in Fig. 6 are centered around $(78\degree,70\degree)$ or $(70\degree,78\degree)$.

From Figs. 4--6, we can conclude that for the stochastic signal model, 1) EM has similar convergence with MEM, and 2) SAGE cannot always outperform EM and MEM.

\begin{figure}[t] \centering
\includegraphics[scale=0.6]{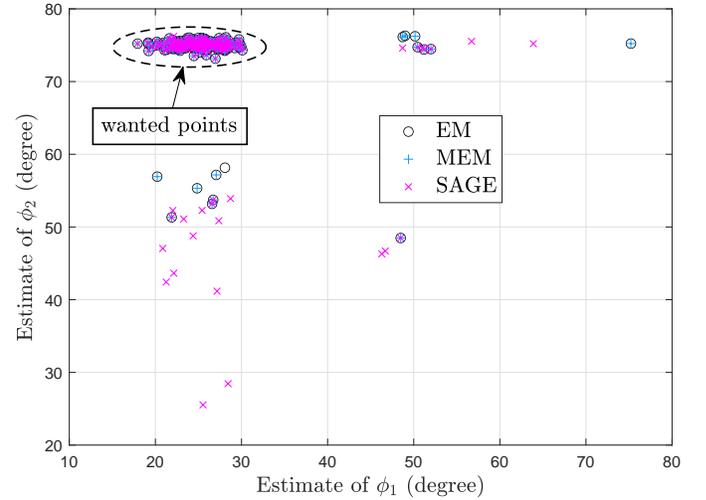}
\vspace{0cm}\caption{Scatter plot of the DOA estimates obtained from EM, MEM, and SAGE for the stochastic signal model under 200 independent realizations with $\phi_1=25\degree$, $\phi_2=75\degree$, $P_1=-4~\mathrm{dB}$, $P_2=2~\mathrm{dB}$, $\sigma=4~\mathrm{dB}$, $\phi^{(0)}_1=40\degree$, $\phi^{(0)}_2=60\degree$, $\mathbf{P}^{(0)}=\mathbf{1}$, $\boldsymbol{\sigma}^{(0)}=[0.5~0.5]^T$, and $\sigma^{(0)}=1$.}\vspace{0cm}
\end{figure}


\begin{figure}[t] \centering
\includegraphics[scale=0.6]{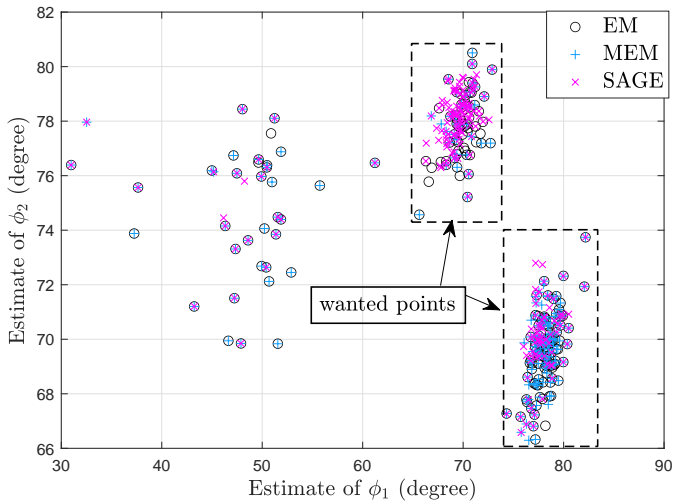}
\vspace{0cm}\caption{Scatter plot of the DOA estimates obtained from EM, MEM, and SAGE for the stochastic signal model under 200 independent realizations with $\phi_1=70\degree$, $\phi_2=78\degree$, $P_1=-2~\mathrm{dB}$, $P_2=-1~\mathrm{dB}$, $\sigma=4~\mathrm{dB}$, $\phi^{(0)}_1=55\degree$, $\phi^{(0)}_2=63\degree$, $\mathbf{P}^{(0)}=\mathbf{1}$, $\boldsymbol{\sigma}^{(0)}=[0.5~0.5]^T$, and $\sigma^{(0)}=1$.}\vspace{0cm}
\end{figure}

\subsection{Deterministic and Stochastic Signal Models}

The EM, MEM, and SAGE algorithms for the deterministic signal model can process samples from the stochastic signal model, which means that the algorithms for the stochastic signal model can be compared to these for the deterministic signal model. The above numerical results have shown that EM has similar convergence with MEM, so we only compare EM and SAGE for both models in this subsection for simplicity.

Since both models have the same DOA parameter $\boldsymbol{\theta}$, the stopping criterion $\Vert\boldsymbol{\theta}^{(k)}-\boldsymbol{\theta}^{(k-1)}\Vert\le\epsilon$ is suitable. Fig. 7 shows a scatter plot of the DOA estimates obtained by EM and SAGE under 50 independent realizations. The same samples of each realization are processed by both algorithms for both models. From Fig. 7, we can observe that both algorithms for both models obtain consistent DOA estimates.

Based on Fig. 7, Fig. 8 compares the numbers of iterations. We can easily observe that EM for the deterministic signal model generally requires a larger number of iterations than EM for the stochastic signal model. Moreover, SAGE for the deterministic signal model generally requires a smaller number of iterations than SAGE for the stochastic signal model. More importantly, SAGE for the deterministic signal model always requires the smallest number of iterations for each realization. Thus, we can conclude that SAGE for the deterministic signal model is superior to the other algorithms in the computational cost.



\begin{figure}[t] \centering
\includegraphics[scale=0.6]{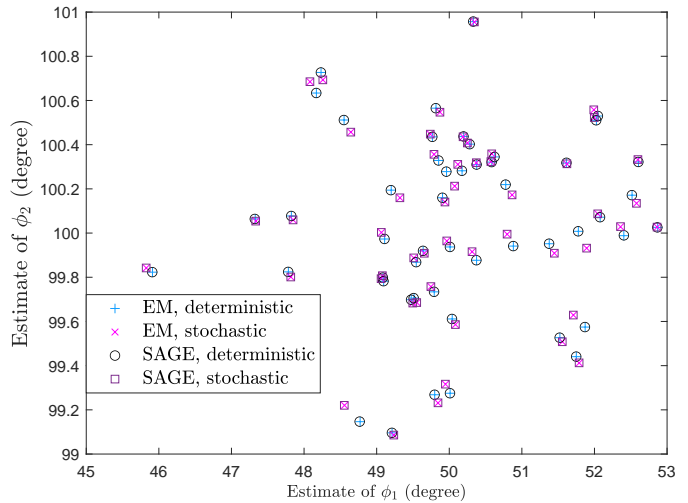}
\vspace{0cm}\caption{Scatter plot of the DOA estimates obtained from EM and SAGE under 50 independent realizations with $\phi_1=50\degree$, $\phi_2=100\degree$, $P_1=-4~\mathrm{dB}$, $P_2=4~\mathrm{dB}$, $\sigma=4~\mathrm{dB}$, $\phi^{(0)}_1=55\degree$, $\phi^{(0)}_2=95\degree$, $\mathbf{S}^{(0)}=[\mathbf{1}~\mathbf{1}]^T$, $\mathbf{P}^{(0)}=\mathbf{1}$, and $\sigma^{(0)}=1$.}\vspace{0cm}
\end{figure}


\begin{figure}[t] \centering
\includegraphics[scale=0.6]{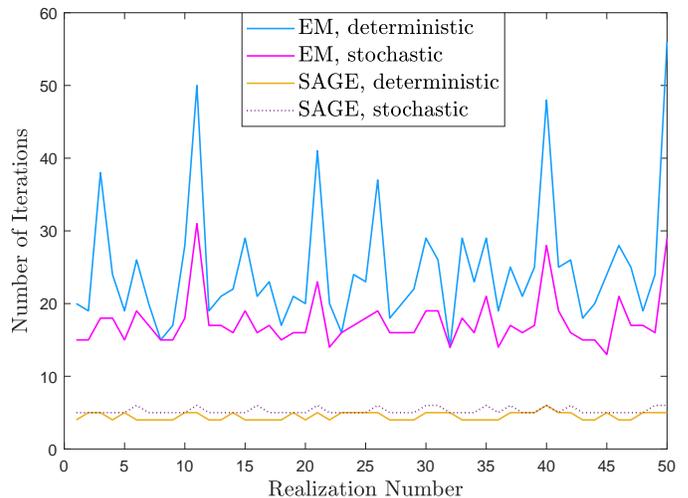}
\vspace{0cm}\caption{Numbers of iterations required by EM and SAGE under 50 independent realizations with $\phi_1=50\degree$, $\phi_2=100\degree$, $P_1=-4~\mathrm{dB}$, $P_2=4~\mathrm{dB}$, $\sigma=4~\mathrm{dB}$, $\phi^{(0)}_1=55\degree$, $\phi^{(0)}_2=95\degree$, $\mathbf{S}^{(0)}=[\mathbf{1}~\mathbf{1}]^T$, $\mathbf{P}^{(0)}=\mathbf{1}$, and $\sigma^{(0)}=1$.}\vspace{0cm}
\end{figure}

\end{subsection}

\section{Conclusion}

We have developed the EM and SAGE algorithms for DOA estimation in unknown uniform noise and proposed an MEM algorithm applicable to the noise assumption. Then, we improve the EM, MEM, and SAGE algorithms to ensure the stability when the powers of sources are unequal. After being improved, numerical results illustrate that the EM algorithm has similar convergence with the MEM algorithm, the SAGE algorithm outperforms the EM and MEM algorithms for the deterministic signal model, and the SAGE algorithm converges faster than the EM and MEM algorithms for the stochastic signal model. In addition, numerical results indicate that when these algorithms process the same samples from the stochastic signal model, the SAGE algorithm for the deterministic signal model requires the fewest iterations.

\begin{IEEEbiographynophoto}{Ming-yan Gong}
received the B.Eng. degree in metal material engineering from the Jiangsu University of Science and Technology, Zhenjiang, China, in 2016
and the M.Eng. degree in signal and information processing from the Nanjing University of Posts and Telecommunications, Nanjing, China, in 2019. He is currently working toward the Ph.D. degree with the Beijing Institute of Technology, Beijing, China. His research interests include array signal processing and MIMO communications.
\end{IEEEbiographynophoto}

\begin{IEEEbiographynophoto}{Bin Lyu}
received the B.E. and Ph.D. degrees from the Nanjing University of Posts and Telecommunications (NJUPT), Nanjing, China, in 2013 and 2018, respectively. He is currently an Associate Professor with NJUPT. His research interests include wireless communications and signal processing.
\end{IEEEbiographynophoto}

\end{document}